\documentclass[12pt,preprint]{aastex}

\newcommand{\lsim}{\mathrel{\hbox{\rlap{\lower.55ex \hbox {$\sim$}}
 \kern-.3em \raise.4ex \hbox{$<$}}}}
\newcommand{\gsim}{\mathrel{\hbox{\rlap{\lower.55ex \hbox {$\sim$}}
 \kern-.3em \raise.4ex \hbox{$>$}}}}
\def\mso{\,\mathrm{M}_\odot}

\def\simle{\mathrel{\hbox{\rlap{\hbox{\lower4pt\hbox{$\sim$}}}\hbox{$<$}}}}
\def\simgr{\mathrel{\hbox{\rlap{\hbox{\lower4pt\hbox{$\sim$}}}\hbox{$>$}}}}

\shorttitle{Collapsars and Cosmic Metallicity Evolution}
\shortauthors{Langer and Norman}

\begin{document}

\title{On the Collapsar Model of Long Gamma-Ray Bursts: Constraints
from Cosmic Metallicity Evolution}

\author{N. Langer\altaffilmark{1} and C.A. Norman\altaffilmark{1,2,3}}


\altaffiltext{1}{Astronomical Institute, Utrecht
University, NL-3584 CC Utrecht, The Netherlands 
}
\altaffiltext{2}{The Johns Hopkins University, Homewood Campus,
Baltimore, MD 21218 
}
\altaffiltext{3}{Space Telescope Science Institute, 3700 San Martin
Drive, Baltimore, MD 21218}

\begin{abstract} 
We explore the consequences of new observational and theoretical
evidence that long gamma-ray bursts prefer low metallicity
environments. Using recently derived mass-metallicity correlations and
the mass function from SDSS studies, and adopting an average
cosmic metallicity evolution from \citet{kewley2005}
and \citet{savaglio2005} we derive expressions for 
the the relative number of massive stars formed below
a given fraction of solar metallicity,
$\varepsilon$, as function of redshift. We demonstrate that about 1/10th
of all stars form with $\varepsilon < 0.1$. Therefore, a picture where
the majority of GRBs form with $\varepsilon < 0.1$
is not inconsistent with an empirical
global SN/GRB ratio of 1/1000.
It implies that (1) GRB's peak at a significantly higher
redshift than supernovae; (2) massive star evolution at low
metallicity may be qualitatively different and; (3) the larger
the low-metallicity bias of GRBs the less likely binary
evolution channels can be significant GRB producers. 
\end{abstract}

\section{Introduction}
It become clear in the last few years that long gamma-ray bursts
are associated with the endpoints of massive star evolution. They occur in star
forming regions at cosmological distances \citep{jakobsson2005}, and 
are assiciated with supernova-type energies. 
The collapsar model explains gamma-ray burst formation via the collapse of
a rapidly rotation massive iron core into a black hole
\citep{woosley1993}. The short time scale of gamma-ray emission
requires a compact stellar size, of the order of lightseconds. This
constraint leaves only massive Wolf-Rayet stars as possible
progenitors.  However, this poses a difficulty: Wolf-Rayet stars in
the local universe are known to have strong stellar winds 
\citep{nugis1998}, which lead to a rapid spin-down \citep{langer1998} --- in
agreement with the absence of signatures of rapid rotation in the
Galactic Wolf-Rayet sample \citep{eenens2004}.

It is the current understanding that the ratio of gamma-ray bursts to supernovae
is about 1/1000, based on about 1 - 2\, 10$^{-6}$ observed
bursts per supernova in the BATSE sample \citep{porciani2001}, and a beaming
factor of $\sim 500$ \citep{frail2001,yonetoku2005}.
This implies that about 1 out of 100 Wolf-Rayet stars
produces a gamma-ray burst \citep{putten2004}. These low values are thought to support the idea
that rather exotic binary evolution channels might constitute the main
evolutionary paths towards gamma-ray bursts \citep{podsiadlowski2004,fryer2005},
corroborated by the two basic problems of single star models 
to produce collapsars \citep{petrovic2005}: (1) the spin-down of the stellar core 
due to magnetic core-envelope coupling \citep{spruit2005,petrovic2005} 
--- which is required to understand
the slow rotation of young pulsars \citep{ott2005} and white dwarfs
\citep{berger2005}; and (2) the spin-down due to Wolf-Rayet winds mentioned above.

However, recent single star models have overcome both problems for
suitable initial conditions. The models by \citet{woosley2005} and
\citet{yoon2005} avoid problem (1) by rapid rotation --- which keeps
the stars nearly chemically homogeneous and thus avoids the formation
of a massive envelope --- and problem (2) by choosing a low enough
metallicity --- which, according to recent evidence \citep{crowther2002, vink2005},
reduces the Wolf-Rayet mass loss rates. 
These are the first single star evolution models fulfilling
the requirements of the collapsar model which are at the same time fully consistent
with the slowly rotating stellar remnants in our Galaxy. However, they
predict long GRBs only for  metallicities of about $Z/Z_{\odot}
\lesssim 0.1$.

In this context, the growing empirical evidence
that the long bursts indeed prefer a low-$Z$ environment is remarkable.
While indirect evidence from gamma-ray
burst host galaxies is pointing towards low metallicity 
\citep{fynbo2003, conselice2005, fruchter2005}, direct metallicity
determinations yield sub-solar values down to 1/100th of the solar
metallicity \citep{gorosabel2005, chen2005, starling2005}.
While the observational and the theoretical evidence for long
gamma-ray bursts occurring at low metallicity needs further confirmation, we
are motivated by the findings reported above to explore the
consequences of such a possibility. 

\section{Method and results}
In order to quantify the amount of star formation at a given
metallicity, we use the Schechter distribution function of galaxy
masses and then substitute in it the well-defined mass-metallicity
relation. We use the mass function and metallicity-mass correlation
determined in comprehensive SDSS studies and deep surveys. For the
mass function there are fine studies by \citet{cole2001},
\citet{bell2004} and \citet{panter2004}, and we use the latter here. For
the mass-metallicity relation we use the classic studies of
\citet{tremonti2005} and \citet{savaglio2005}. From Equation (1) of
\citet{panter2004}
 
\begin{equation}
\label{massfunctioneqn}
\Phi (M) = \Phi_* \left( M\over M_* \right)^{\alpha} e^{(-M/M_*) },
\end{equation} 
where $\alpha = -1.16$ and $\Phi_* = 7.8 \times 10^{-3}h^3 Mpc^{-3}$.
The fraction $\Psi(M)$ of the mass density in galaxies with a mass
less than $M$ is then
\begin{eqnarray}
\label{yieldeqn}
\Psi(M) & = & { \int_0^M M \Phi(M) dM \over \int_0^{\infty} M \Phi(M) dM} 
  \nonumber \\
 & = & {\hat\Gamma(\alpha+2, (M/M_*)) \over \Gamma(\alpha +2)} ,
\end{eqnarray}
where $\hat\Gamma$ and $\Gamma$ are the incomplete and complete gamma
function. We then choose the galaxy mass-metallicity relation of the
form: $M/M_* = K (Z/Z_{\odot})^{\beta}$ where $K$ and $\beta$ are
constants. For simplicity we choose to use the linear bisector fit to
the mass metallicity relation derived by \citet{savaglio2005} at redshift,
$z=0.7$. This is parallel to the quadratic fit of \citet{tremonti2005} at
low masses and metallicities that we consider here. The form we derive
is very close to $\beta=2$ and $K=1$ and we assume these values. This gives
a pleasing Gaussian form to the metallicity function which may be of
more general applicability. We now derive the fractional mass density
belonging to metallicities below metallicity $Z$ at a given redshift $z$ as
\begin{equation}
\label{yieldevolutioneqn}
\Psi \left(Z\over Z_{\odot}\right) = 
      {\hat\Gamma(\alpha+2,(Z/Z_{\odot})^{\beta} 10^{0.15\beta z}) \over
           \Gamma(\alpha+2)} ,
\end{equation}
where we used the average cosmic metallicity scaling as ${d [Z] \over
dz} = -0.15$ dex per unit redshift from Kewley \& Kobulnicky (2005, 2006).  We used this scaling since the mean metallicity is given by:
\begin{eqnarray}
\label{meanmetallicityeqn}
\langle Z \rangle & = & { \int_0^\infty Z(M) M\Phi(M) dM \over \int_0^{\infty} M \Phi(M) dM} 
    \nonumber \\
 & = & K^{-1 \over \beta} Z_{\odot}{\Gamma(2 +\alpha + {1 \over \beta})) \over \Gamma(\alpha +2)} ,   
\end{eqnarray}
and thus the mean metallicity is linearly proportional to $Z_{\odot}$,
$\langle Z \rangle = constant \times Z_{\odot}$.  The ratio of the
gamma-functions in the above equation gives $0.80$ for $\beta =5$,
and $0.73$ for $\beta=2$, assuming $\alpha = -1.16$ in both cases.
Our scaling simply gives self-consistently $\langle Z \rangle =
Z_{\odot} 10^{-\gamma z}$ where from Kewley and Kobulnicky(2005, 2006)
we have $\gamma =0.15$ that we adopt here. The $\gamma$ value found by
\citet{savaglio2005} is approximately $\gamma \sim 0.3-0.4$ (S.Savaglio,
private communication).

This ansatz includes various simplifications. Because the parameter we
need here is a ratio of two moments of the mass function the evolution
of the normalisation of the mass function (the cosmic stellar mass
density) cancels. The remaining evolutionary term is essentially the
mean cosmic metallicity evolution. Since we are dealing here with
massive star formation from the interstellar medium and since the
metallicity constraint is from the physics of line-driven winds we use
the scaling with redshift derived from emission-line studies by Kewley
and Kobulnicky (2005, 2006) for [O/H].  The study of
\citet{gallazzi2005} derives stellar metallicities and gives in their
Figure 6 the distribution of stellar metallicity in galaxies at low
redshift. As they note, in general, the stellar metallicity is always
lower than the gas-phase metallicity and our use of Kewley and
Kobulnicky's gas-phase evolution is conservative. We are assuming in
these star-forming galaxies that the evolving metallicity distribution
we have derived also applies to the ISM in which the stars are
forming.

 For our fiducial numbers $\alpha
=-1.16$,  $\beta =2$ and $K=1$we thus obtain
\begin{equation}
\label{yieldnumericaleqn}
\Psi \left(Z\over Z_{\odot}\right) = 
      {\hat\Gamma(0.84,(Z/Z_{\odot})^{2} 10^{0.30 z)})\over \Gamma(0.84)},
\end{equation}
where $\Gamma(0.84) \simeq 1.122$. 

\begin{figure}[t]
\begin{center}
\resizebox{0.80\hsize}{!}{\includegraphics[angle=270]{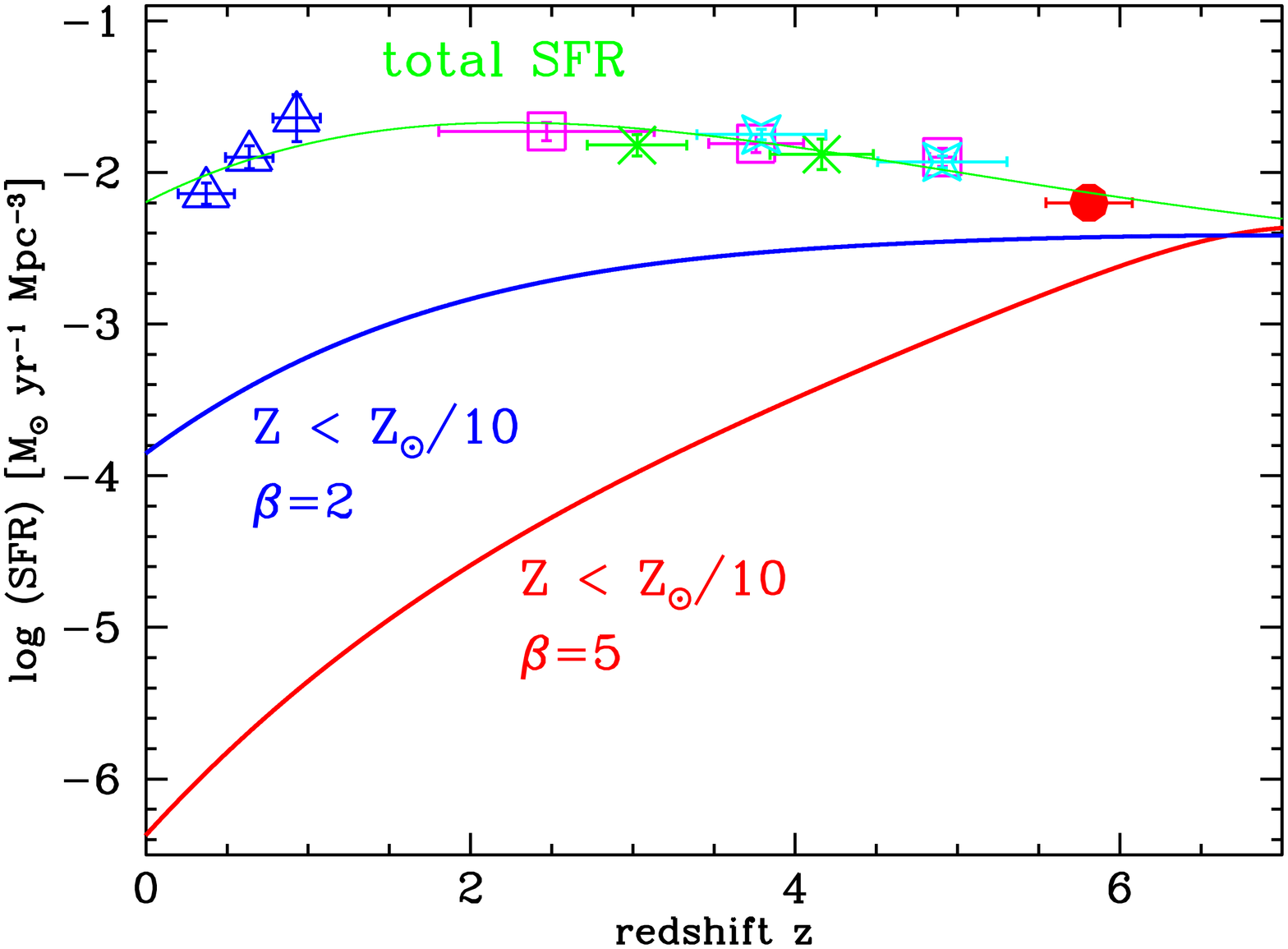}}
\caption{ Formation rate of stars with a metallicity below
$Z_{\odot}/10$, ($\varepsilon = 0.1$) for a fiducial
galaxy mass-metallicity relation $M\sim Z^{\beta}$ for $\beta=2$ (blue
curve) and  $\beta=5$ (red
curve).  These curves have been obtained by multiplication of the fit
to the total star formation rate as function of redshift (weak green
line) obtained from the empirical data presented by Bouwens et
al. (2004; points with error bars) with the fraction $\Psi$ of stars
born with a metallicity below a specified value. \label{SFR} }
\end{center}
\end{figure}

The result of folding $\Psi$ 
with the total star formation rate history 
is shown in Figure~\ref{SFR}. Here, the total star
formation rate $r_{\rm SFR}(z)$ is derived from a 4th-order polynomial fit to the data
presented in \citet{bouwens2004}.

In order to be able to compare with observations, we convolve the
fraction $\Psi$ of stars born with a metallicity of less than
$Z_{\odot}\varepsilon$ with the comoving volume element of the
Friedman-Robertson-Walker metric. The quantity $d_m(z)$ used below is
the proper motion distance for our cosmological parameters. We are
using the standard cosmological parameters of $H_0=
70\,$km$\,$s$^{-1}\,$Mpc$^{-1}$, $\Omega_{\rm M}=0.3$, and
$\Omega_{\Lambda}=0.7$.

\begin{figure}[t]
\begin{center}
\resizebox{0.80\hsize}{!}{\includegraphics[angle=270]{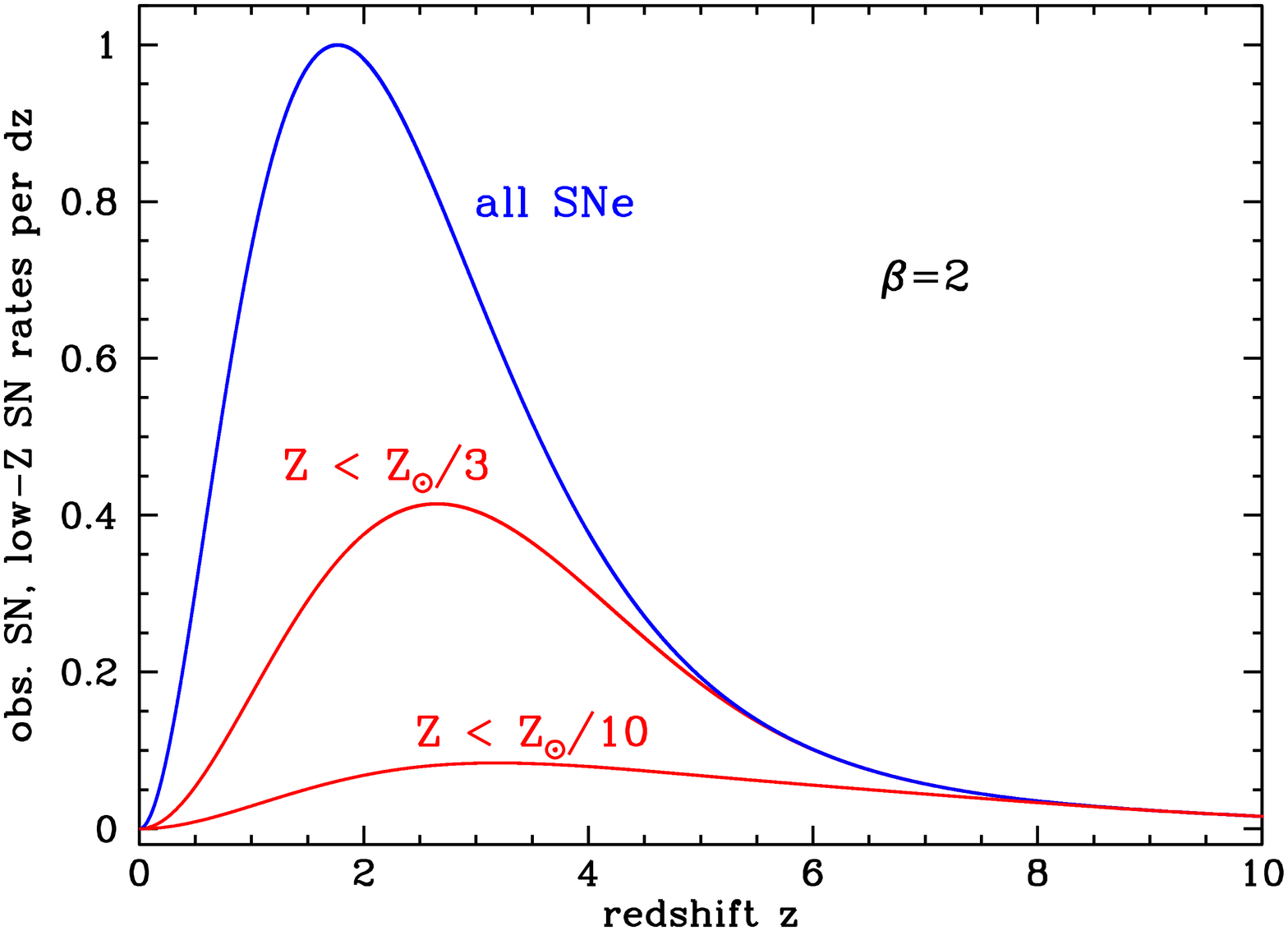}}
\hfill
\resizebox{0.80\hsize}{!}{\includegraphics[angle=270]{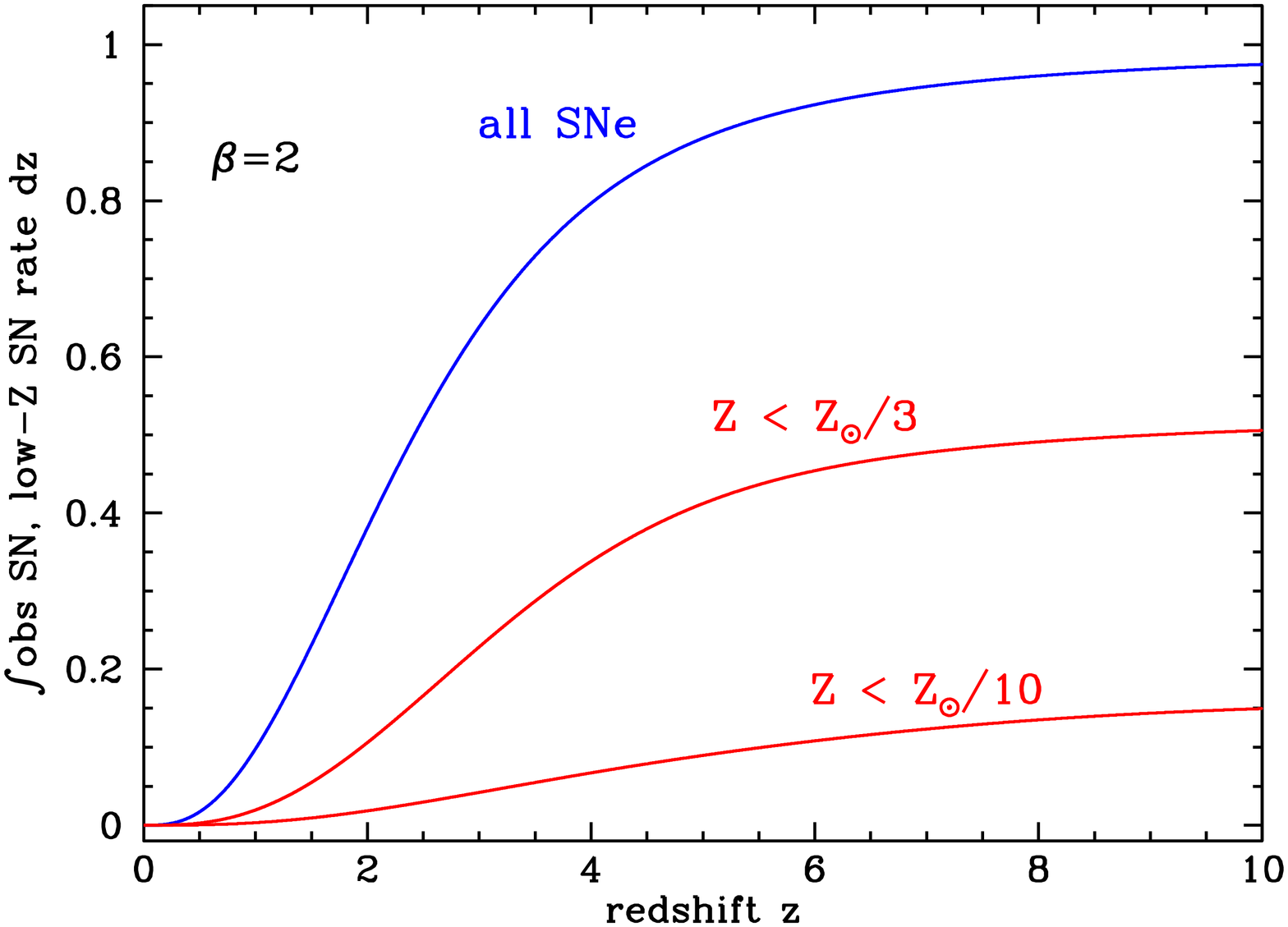}}
\caption{Normalised differential and cumulative core collapse 
 supernovae rates as function of redshift as preceived by an 
 unbiased observer (cf. Eq.~6), for all stars,
 and for stars with metallicities less than $Z_{\odot}/3$
 and $Z_{\odot}/10$.
  \label{SNrates}}
\end{center}
\end{figure}

Including the time dilation effect due to redshift that slows the
rates with redshift, the unbiased observed rate of core collapse
supernovae from stars with a metallicity below $Z_{\odot}\varepsilon$
then becomes
\begin{equation}
\label{corecollapseSNrate}
R_{\rm SN}(\varepsilon) = f_{\rm SN} 4\pi \int_0^{z_{\rm max}}
    {\Psi(z,\varepsilon) r_{\rm SFR}(z) \over z+1} d^2_{\rm M}(z)
    \left( d(d_{\rm M}(z)) \over dz \right),
\end{equation} 
where
\begin{equation}
\label{corecollapseunitstareqn}
f_{\rm SN} = \int_{8 M_{\odot}}^{100 M_{\odot}} a \varphi (m) dm  \simeq 0.0074
\end{equation}
is the number of core collapse events per solar mass of stars formed, where
$a = 1 M_{\odot} / \int \varphi(m) dm$, $\varphi(m)$ is the Salpeter initial
stellar mass function, and $f_{\rm SN}$ is the number
of core collapse events per solar mass of star formation.
Figure~\ref{SNrates} shows the resulting supernova rates,
and their integrated value, for various values of $\varepsilon$. 



In Table~\ref{Tableone}, we derive the preceived global and local ratios of 
low-metallicity to all supernovae, for various metallicity thresholds
and for two galaxy mass-metallicity exponents.  From these,
we compute formation rates of low-metallicity black holes by
assuming supernovae to form from stars above 8$\mso$,
but black holes to form only from stars above 30$\mso$,
which --- for a Salpeter IMF --- gives a BH/SN ratio of 0.14.
Since black hole formation is only one out of three criteria for
GRB production within the collapsar frame, our derived numbers
give an upper limit to the GRB production rate from low metallicities,
independent of the stellar evolution scenario considered. For
$\varepsilon = 0.1$, this corresponds to less than 22 GRBs per
1000 SNe globally in the universe, and to less than 3/1000 at low redshift. 
This is what stars with a metallicity of $Z_{\odot}/10$ and below
can provide maximally. Table~1 shows further that the constraint of
achieving 1 GRB per 1000 SNe globally in the universe results
in about 1 GRB per 10$\,$000 SNe in the local universe.

We note that these numbers seem to be remarkably insensitive to details
of the star formation history. 
The last row in Table~\ref{Tableone} is computed by
using the star formation history shown by \cite{firmani2005},
which is similar to the one used here up to redshift~2, but which
remains at the top level until about $z=6$ (cf. their Fig.~2).
However, Table~1 also shows that the redshift at which the GRB rate peaks
does depend on the star formation history and can thus not yet be predicted
reliably. 


\section{Discussion and conclusions}

To produce a GRB, Wolf-Rayet stars at core collapse are required to
have sufficient angular momentum. Stellar
evolution models which include magnetic fields predict 
too slowly-rotating cores for models which
develop an extended, massive envelope after the main sequence.
Current evolutionary models that include rotation 
predict extended envelopes for the vast
majority of massive stars in the Galaxy or Magellanic Clouds 
--- in agreement with the number of blue and
red supergiants \citep{maeder2001, maeder2005}. Only the fastest
rotators are thought to be able to avoid extended envelopes, for
metallicities below about $Z_{\odot}/10$, or $\varepsilon = 0.1$
\citep{yoon2005, woosley2005}. 

Most importantly, the numbers derived for $\varepsilon = 0.1$ in Table~1 appear not 
to be in conflict with observations. Per 1000 supernovae in the universe,
160 are predicted to occur from stars with $Z > Z_{\odot}/10$, out of which
22 would produce a black hole.  Thus, producing
one GRB per 1000 supernovae globally in the universe (cf. Sect.~1) seems possible.  

On the other hand, a large fraction of the 22 black holes may be born
without producing a GRB: not all of them may occur in WR stars but rather in more
extended stars \citep{maeder2005}, and the most massive ones 
would lose too much angular momentum
in a wind, even for metallicities as low as $Z > Z_{\odot}/10$ \citep{yoon2005}.
While within the chemically-homogeneous-evolution scenario for GRB formation
\citep{yoon2005, woosley2005} a GRB/BH fraction of 1/20 can certainly be obtained
\citep{yoon2006}, this is therefore unlikely 
for exotic binary channels for GRB production 
--- i.e. for channels through which only a small fraction of stars of any initial mass 
evolves.
Clearly, the more the long GRBs are confined to low metallicities, the more
unlikely it is that binary evolution is needed to explain the majority of events. 


The empiric cosmic GRB to SN ratio of about 1/1000 (cf. Sect.~1) can not directly 
rule out more extreme values of $\varepsilon$, i.e. $\varepsilon= 0.1 -
0.01$ (cf. Table~1); in fact, for $\varepsilon= 0.01$, about
10\% of all massive stars with $Z < Z_{\odot}/100$ would need to produce 
a GRB. However, it would imply that the formation of {\em every} black hole 
would be accompanied by a GRB. Furthermore, 
the GRB rate would peak only at a redshift of about $z=10$.
A value of $\varepsilon= 0.01$ appears thus unlikely.
Furthermore, our models with $\beta=5$ produce such a small local
GRB/SN ratio that they seem to be ruled out.

We find that a restriction of GRBs to low metallicities ($Z < Z_{\odot}/10$, i.e. $\varepsilon= 0.1$)
has the following consequences:

\begin{itemize}
\item GRBs do not follow star formation in an unbiased manner.
  For example, for an overall star formation rate which predicts a preceived SN peak at a redshift
  of $z_{\rm SM} \simeq 1.8$
  we find, for $\varepsilon =0.1$, that the GRB rate peaks at a redshift
  of $z_{\rm GRB}\simeq 3.2$ (Fig.~2 and Tab.~1; see also \citet{firmani2005}
\item Local massive galaxies, like our Milkyway, are not expected to host long GRBs.
  The last long GRB in our Galaxy should have occurred several gigayears ago.
\item The global and local GRB to SN ratios appear to be insensitive to the 
  details of the cosmic star formation history, while
  the redshift of the peak GRB rate can vary appreciably (Tab.~1).  
\item The local GRB/core-collapse ratio is much smaller than
  the one obtained from averaging over the universe; i.e., by
  one order of magnitude for $\varepsilon =0.1$ (Tab.~1).
\item
  We obtain the expected result that the number of massive stars
  in the universe with a metallicity below a critical value $\varepsilon$
  does roughly scale with $\varepsilon$. I.e., For $\varepsilon =0.1$, 
  we find a ratio of low-metallicity
  ($Z<Z_{\odot}/10$) to total global supernova rates of 0.16.
  Further, the probability that a randomly chosen burst
  has a metallicity of $Z_{\odot}/100$ is about 10\%.
\item We derive the most likely redshift for gamma-ray bursts of specified
  metallicity for unbiased observations. For example, for a metallicity of
  $Z_{\odot}/100$ -- as found for GRB050730 --, we find GRBs to occur
  most likely at redshifts of $z > 6$. Locally, the ratio of GRBs with a
  metallicity of $Z_{\odot}/100$ to all GRBs is about 0.02 (Tab.~1).
\item The larger the low-metallicity bias of long GRBs, the less likely
  can binary scenarios explain the major fraction of them.
\item A confirmation of the low-metallicity bias of long GRBs to values
  of the order of $\varepsilon =0.1$ would imply that fast rotation 
  may be much more common at low metallicity among massive stars.
\end{itemize}

\begin{acknowledgements}
We are  grateful to Bram Achterberg, Andrew Fruchter, Karl Glazebrook,
Max Petini, Sandra Savaglio, Rosemary Wyse and Stan Woosley for 
very interesting and useful discussions.
C.N. is very pleased to thank the Astronomical Institute in Utrecht
for stimulating hospitality.
\end{acknowledgements}

\clearpage


\clearpage

\clearpage

\clearpage
\begin{table*}[th]
\begin{center}
\caption{For various upper metallicity limits $\varepsilon$, and
galaxy mass-metallicity exponents $\beta$: 
the number of supernovae with $Z/Z_{\odot} < \varepsilon$ (or ``low-$Z$ supernovae'') per
1000 supernovae in the universe as seen from an unbiased observer 
(cf. Eq.~6, and Fig.~2), the corresponding number
of ``low-$Z$ black holes'' per 1000 supernovae, both numbers at redshift $z=0$, the local
($z=0$) number of GRBs per 1000 supernovae required to have a global GRB/SN ratio of 1/1000,
and the redshifts at which the SN and GRB-rates are preceived as maximum by an unbiased observer (cf. Fig.~2). 
For the last row, a different underlying star formation history has been assumed (see text).
\label{Tableone}}
\vspace{0.1cm}
\begin{tabular}{c c c c c c c c c}
\hline
$\varepsilon$ & $\beta$ &
$\langle {{\rm low} Z\, {\rm SNe} \over 1000\, {\rm SNe}} \rangle$ & $\langle {{\rm low} Z\, {\rm BHs} \over 1000\, {\rm SNe}} \rangle$ & 
$\left( {{\rm low} Z\, {\rm SNe} \over 1000\, {\rm SNe}} \right)_{z=0}$ & $\left( {\rm low} Z\, {\rm BHs} \over 1000\, {\rm SNe}\right)_{z=0}$ & 
$\left( {\rm GRB} \over 1000\, {\rm SNe} \right)_{z=0}$ & $z_{\rm SN}$ & $z_{\rm GRB}$ \\
\hline
0.3 & 2 & 520 & 73 & 130 & 18 & 0.25 & 1.8 & 2.7 \\ 
{\bf 0.1} & {\bf 2} & {\bf 160}& {\bf 22}& {\bf 20} & {\bf 3}& {\bf 0.13} & " & {\bf 3.2} \\
0.03& 2 & 34 & 5 & 3 & 0.4 & 0.09 & " & 5.3 \\
0.01& 2& 7 & 1 & 0.5 & 0.07 & 0.08 & " & 9.8 \\
0.3 & 5 & 380 & 53 & 7 & 1 & 0.02 & " & 3.7 \\
0.1 & 5 & 60 & 8 & 0.007 & 0.001 & 0.0001 & " & 7.0 \\
\hline
0.1 & 2 & 160 & 22 & 20 & 3 & 0.13 & 2.0 & 5.3 \\
\hline
\end{tabular}
\end{center}
\end{table*}


\end{document}